
%
\magnification=\magstep1
\openup 2\jot

\font\list=cmcsc10


\newcount\refno
\refno=0
\def\nref#1\par{\advance\refno by1\item{[\the\refno]~}#1}
\def\npb#1 #2 #3.{{\it Nucl.\ Phys.\ \rm B\bf#1}, #2 (#3).}
\def\plb#1 #2 #3.{{\it Phys.\ Lett.\ \rm B\bf#1}, #2 (#3).}
\def\prd#1 #2 #3.{{\it Phys.\ Rev.\ \rm D\bf#1}, #2 (#3).}

\overfullrule=0pt       

\def\half{\textstyle{1\over2}}

\def\spose#1{\hbox to 0pt{#1\hss}}
\def\lta{\mathrel{\spose{\lower 3pt\hbox{$\mathchar"218$}}
     \raise 2.0pt\hbox{$\mathchar"13C$}}}

\hbox{ }
\rightline {DTP/94/51}
\rightline {DAMTP/R-94/59}
\rightline {hep-ph/9412232}
\rightline {December 1994}
\vskip 0.8truecm

\centerline{\bf A NOTE ON THE FINITENESS OF CURVATURE AT COSMIC STRING CUSPS}
\vskip 0.8truecm

\centerline{Ruth Gregory\footnote{$^\spadesuit$}{\sl Email:
rg10012@amtp.cam.ac.uk}}
 \vskip 2mm
\centerline{ \it Centre for Particle Theory, Department of Mathematical
Sciences}
\centerline{\it University of Durham, Durham, DH1 3LE, U.K.}
\vskip 1mm
\centerline{\it and}
\vskip 1mm
\centerline{ \it D.A.M.T.P. University of Cambridge}
\centerline{ \it Silver Street, Cambridge, CB3 9EW, U.K.}

\vskip 1.0cm
\centerline{\list abstract}
\vskip 3mm

{ \leftskip 5truemm \rightskip 5truemm

\openup -1\jot

We examine a Nambu-Goto string trajectory in the neighbourhood of a cusp and
determine the extrinsic curvature invariants. These are demonstrated to be
finite, in contradiction with naive expectation. Thus the Nambu action is a
valid approximation for the motion of a cosmic string vortex even at cusps.
This suggests that cusp annihilation is not a feasible mechanism
for particle radiation from cosmic strings.

\openup 1\jot


}

\vfill\eject
\footline={\hss\tenrm\folio\hss}

The motion of a cosmic string (see [1] for a recent review), determined
by the Nambu action, generically
contains cusps, or points at which the string instantaneously reaches the speed
of light. The Nambu action is the lowest order term in an effective action for
the cosmic string motion[2], and is valid when the string width, w, is much
less
than its radius of curvature, $K^{-1}$. It has generally been thought that this
approximation breaks down at a cusp[3], where the extrinsic curvature of the
worldsheet was presumed infinite. In this note, we point out that this is not
the case.

Our starting point is the Nambu-Goto action. Let $X^\mu$ be the spacetime
coordinates of the string worldsheet, and $\sigma^A=\{\sigma,\tau\}$, be a set
of worldsheet coordinates, with $\tau$ understood as being a timelike
coordinate. Then the induced metric on the worldsheet is given by
$$
\gamma_{AB} = {\partial X^\mu \over \partial \sigma^A}
{\partial X^\nu \over \partial \sigma^B} \eta_{\mu\nu}
\eqno (1)
$$
where $\eta_{\mu\nu} = $Diag$(c^2, -1, -1, -1)$ is the Minkowski spacetime
metric. The action is proportional to the area of the surface swept out
by the string, and is given in terms of its intrinsic metric by
$$
S = - {\mu\over c} \int \sqrt{-\gamma} d^2\sigma = -{\mu\over c} \int d\sigma
d\tau \{ ({\dot X}^\mu X'_\mu)^2 - {\dot X}^{\mu^2} X^{\prime\mu ^2} \}
\eqno (2)
$$
where $'$ denotes $\partial/\partial\sigma$ and $\cdot$  denotes
$\partial/\partial\tau$. $\mu$ is the energy per unit
length of the string and is of order $10^{38}$Jm$^{-1}$ for a GUT string.
Henceforth we will set $\hbar=c=1$.

There is a great deal of freedom in the choice of coordinates on the string
worldsheet, so following Kibble and Turok[4], we fix our  gauge by setting
$\tau=t$, the Minkowski time coordinate, and imposing the conformal gauge
conditions
$$
{\dot X}^\mu X'_\mu = 0 \;\;\; ; \;\;\;\;\;\;\;\;\;\;\;\;\; {\dot
X}^{\mu^2} + X^{\prime\mu ^2} = 0
\eqno (3)
$$
so that the metric is conformally flat ($\gamma_{AB}\propto\eta_{AB}$).
Writing the worldsheet coordinates  in the form
$$
X^\mu = (\tau, {\bf r}(\sigma,\tau)) \ ,
\eqno (4)
$$
the equations of motion are written in terms of {\bf r} as
$$
\eqalignno{
{\dot {\bf r}} . {\bf r}' & = 0 & (5a) \cr
{\dot{\bf r}}^2 + {{\bf r}'}^2 &=1 & (5b) \cr
{\ddot{\bf r}} - {\bf r}'' &=0 & (5c) \cr
}
$$
The first two of these equations say that ${\dot{\bf r}}$ is the observable
velocity of the string perpendicular to itself, and  that the parameter
$\sigma$ (which satisfies $d\sigma = dr/\sqrt{1-{\dot{\bf r}}^2}$) measures
equal energy intervals along the string.  (5c), the
equation of motion, is a simple wave equation, and hence the solution for {\bf
r} consists of a superposition of arbitrary left and right moving waves:
$$
{\bf r} = \half [ {\bf a} ( \sigma - \tau ) + {\bf b} (\sigma+\tau) ]
\eqno (6)
$$
with the constraints
$$
{{\bf a}'}^2 = {{\bf b}'}^2 = 1\ .
\eqno (7)
$$
Therefore ${{\bf a}'}$ and ${\bf b}'$ lie on a unit sphere, but are otherwise
arbitrary. If we are dealing with a closed loop of length $L$, then both {\bf
a}
and {\bf b} are periodic with length $L$, but the string motion actually has
periodicity $L/2$, since ${\bf r}(\sigma+L/2,\tau+L/2) = {\bf r}(\sigma,\tau)$.
Additionally, both ${\bf a}'$ and ${\bf b}'$ must average to zero in the centre
of mass frame since $\int_0^L d\sigma {\bf r}'=\int_0^Ld\sigma{\dot{\bf r}}=0$.

Returning to (1), and inputting (4) and (7) gives
$$
\gamma_{AB} = \half(1 + {\bf a}'.{\bf b}')\eta_{AB}
\eqno (8)
$$
in conformal form as required. Note that
if ${\bf a}' = -{\bf b}'$, ${\bf a}'.{\bf b}'=-1$, and $\gamma_{AB}$
becomes degenerate, this reflects the fact that the worldsheet is null
and is called a cusp. At a cusp ${\dot{\bf r}}=1, {\bf r}'=0$,
i.e.~the string is travelling at the speed of light, and $\sigma$ is varying
arbitrarily rapidly at this point, so the mass concentrated there is
technically infinite. It is fortunate that the cusp has measure zero!
Nonetheless, ultra-relativistic velocities, and high density concentrations
could indicate a breakdown of the approximation used to obtain the motion of
the cosmic string, which after all is a finite thickness vortex; on the other
hand, these features could be an artefact of the nullity of the worldsheet, and
provided that this occurs only at an isolated point there may not be a problem.
In order to investigate this, we examine the extrinsic curvature near a cusp.

A worldsheet living in four dimensions has co-dimension 2, and thus has 2
families of normals, $n_{_i}^\mu$, satisfying
$$
n_{_i}^\mu n_{_j}^\nu \eta_{\mu\nu} = -\delta_{ij} \;\;\; ; \;\;\;\;\;\;
n_{_i}^\mu {\partial X^\mu \over\partial \sigma^A} = 0
\eqno (9)
$$
and two associated extrinsic curvatures
$$
K_{_iAB} = {\partial X^\mu \over\partial \sigma^A}
{\partial X^\nu \over\partial \sigma^B} \nabla_{(\mu} n_{_i\nu)}\ ,
\eqno (10)
$$
(see [5] for greater detail on this formalism.)
There is of course an SO(2) gauge freedom in the choice of normals, $n_{_i}^\mu
\to R_{ij} n_{_j}^\mu$, but this will not be relevant in our calculation. An
example of two normals is:
$$
\eqalign{
n_{_1}^\mu &= (1 - ({\bf a}'.{\bf b}')^2)^{-\half} \Bigl ( 1-{\bf a}'.{\bf
b}', {\bf b}' - {\bf a}' \Bigr ) \cr
n_{_2}^\mu &= (1 - ({\bf a}'.{\bf b}')^2)^{-\half} \Bigl ( 0, {\bf
a}'\wedge{\bf b}' \Bigr ) \cr
}
\eqno (11)
$$
note that as we approach a cusp, $n_{_1}^\mu$ tilts over towards the light cone
whereas $n_{_2}^\mu$ flips direction across the cusp, corresponding to our
notion of the light-speed, sharp point of a cusp.

Calculating the associated extrinsic curvatures gives
$$
\eqalign{
K_{_1AB} &= \half (1 - ({\bf a}'.{\bf b}')^2)^{-\half} \left [ (\delta^\tau_A
\delta^\tau_B + \delta^\sigma_A \delta^\sigma_B)({\bf a}''.{\bf b}' - {\bf b}''
. {\bf a}') - 2 \delta^\tau_{(A} \delta^\sigma_{B)} ({\bf a}''.{\bf b}' + {\bf
b}''.{\bf a}') \right ] \cr
K_{_2AB} &= \half (1 - ({\bf a}'.{\bf b}')^2)^{-\half} \left [ (\delta^\tau_A
\delta^\tau_B + \delta^\sigma_A \delta^\sigma_B)({\bf a}'' + {\bf b}''). {\bf
a}'\wedge{\bf b}' + 2 \delta^\tau_{(A} \delta^\sigma_{B)}({\bf a}''- {\bf
b}'').
{\bf a}'\wedge{\bf b}' \right ] \cr
}
\eqno (12)
$$
note $\gamma^{AB} K_{_iAB}\equiv0$ as required by the Nambu equations of
motion.

In order to examine whether a cusp is a singular point for the extrinsic
curvature, we must examine scalars constructed from the curvature. Suitable
worldsheet scalars are
$$
\eqalign{
\Sigma_1 &= K_{_1AB}K_{_1}^{AB} = (1 - ({\bf a}'.{\bf b}')^2)^{-1} [  ({\bf
a}''.{\bf b}')^2 + ({\bf a}'.{\bf b}'')^2 ] \cr
\Sigma_2 &= K_{_2AB}K_{_2}^{AB} = (1 - ({\bf a}'.{\bf b}')^2)^{-1} [
({\bf a}''.{\bf a}'\wedge{\bf b}')^2 + ({\bf b}''.{\bf a}'\wedge{\bf b}')^2
]\cr
\Sigma_3 &= K_{_1AB}K_{_2}^{AB} = (1 - ({\bf a}'.{\bf b}')^2)^{-1} [({\bf a}''.
{\bf b}')({\bf a}''.{\bf a}'\wedge{\bf b}') - ({\bf a}'.{\bf b}'')
({\bf b}''.{\bf a}'\wedge{\bf b}') ] \cr
}
\eqno (13)
$$
These  of
course  depend on the normals, to remove this dependence, one can consider
the combinations $K_{_iAB}K_{_j}^{AB}\delta^{ij}= \Sigma_1+\Sigma_2$ and
$K_{_iAB}K_{_j}^{AB}\epsilon^{ij}= 2\Sigma_3$. The first of these is
proportional to the intrinsic curvature of the worldsheet by virtue of the
Nambu equations implying $K_{_i}-0$, the second quantity is related to the
curvature of the normal bundle of the worldsheet. We choose to consider the
three quantities in (13) separately to ensure that all possible {\it
worldsheet} scalars vanish.

Now consider a neighbourhood of the cusp. Without
loss of generality, let the intersection of ${\bf a}'$ and $-{\bf b}'$ occur at
$(1,0,0)$, i.e.~$\theta = \pi/2, \phi = 0$ on the unit sphere. Also, without
loss of generality, let
$$
\eqalign{
{\bf a(s)}' &= ( \cos\phi(s),\sin\phi(s),0 ) \cr
{\bf b(s)}' &= ( \cos\phi(s), \cos\psi\sin\phi(s), \sin\psi\sin\phi(s)) \cr }
\eqno (14)
$$
in a neighbourhood of the cusp,
where $\psi$ is the angle between ${\bf a}''$ and $-{\bf b}''$, $\psi\neq
0,\pi$.

Then
$$
\eqalign{
-{\bf a}'.{\bf b }' &= \cos \phi_+ \cos\phi_- + \cos\psi\sin\phi_+\sin\phi_-
\cr
{\bf a}'' &= \phi_-' ( -\sin\phi_-,\cos\phi_-, 0)\cr
{\bf b}'' &= \phi_+' ( -\sin\phi_+, \cos\psi\cos\phi_+, \sin\psi\cos\phi_+)
,\cr
}
\eqno (15)
$$
where we have abbreviated $\phi^{(\prime)}(\sigma\pm\tau)$ as
$\phi^{(\prime)}_\pm$. For small $\phi_\pm$, the three scalars in (13) are
given
by
$$
\eqalign{
\Sigma_1 &= { {\phi'}_-^2[\phi_--\cos\psi\phi_+]^2 +
{\phi'}_+^2[\phi_+-\cos\psi\phi_-]^2 \over \phi_+^2+ \phi_-^2 - 2
\cos\psi\phi_+\phi_-} \cr
\Sigma_2 &= { {\phi'}_-^2\sin^2\psi\phi_+^2 +
{\phi'}_+^2\sin^2\psi\phi_-^2 \over \phi_+^2+ \phi_-^2 - 2
\cos\psi\phi_+\phi_-} \cr
\Sigma_3 &= { -{\phi'}_-^2\sin\psi\phi_+[\phi_--\cos\psi\phi_+] -
{\phi'}_+^2\sin\psi\phi_-[\phi_+-\cos\psi\phi_-] \over \phi_+^2+ \phi_-^2 - 2
\cos\psi\phi_+\phi_-} \cr
}
\eqno (16)
$$
Now, note that $\Sigma_1+\Sigma_2 = {\phi'}_-^2 + {\phi'}_+^2$,
and since $[\phi_\mp - \cos\psi\phi_\pm]^2 < \phi_+^2 + \phi_-^2 -
2\cos\psi\phi_+\phi_-$, we have that $\Sigma_1 < {\phi'}_-^2 +
{\phi'}_+^2$. Hence both $\Sigma_1$ and $\Sigma_2$ are generically
bounded at and near the cusp. Similarly, $\Sigma_1+\Sigma_2
+2\Sigma_3 < 2(\phi_-^{\prime 2} + \phi_+^{\prime 2})$, and hence
$\Sigma_3$ is also genericaclly bounded at and near the cusp.

Thus we see that all of the worldsheet scalars that can be formed from the
extrinsic curvature are bounded in the vicinity of a cusp. We are therefore
forced to conclude that a cusp is analogous to a coordinate singularity as far
as the worldsheet curvature is concerned, and does not
represent in any way a breakdown of the approximation in which we treat a
cosmic
string as a Nambu-Goto string. How can this be, for at a cusp the string
appears to double back on itself? The answer lies in what we mean by string
width and string curvature. The width of the string, $w$, is measured in its
local rest frame, and to order $w/K$ remains constant in that frame[5,6].
Moreover
the actual field theoretic vortex solution is, again to order $w/K$, the static
zero curvature vortex solution (such as the Nielsen-Olesen solution for an
abelian Higgs vortex) in the planes normal to the worldsheet[5,6]. The string
curvature, again, is measured in a local rest frame, both normals being
perpendicular to the four-velocity of the worldsheet. Therefore, both string
width and string curvature, being measured in local rest frames, are inherently
four-dimensional quantities, and it is therefore not only misleading, but
wrong,
to take a ``snapshot'' of the string as an external observer and draw
conclusions as to its curvature. The combination of Fitzgerald contractions and
Lorentz time dilations allow the worldsheet to evade any true curvature
singularities at a cusp.

Finally, we would like to remark on kinks. A kink corresponds to a
discontinuity in ${\bf a}'$ and/or ${\bf b}'$, so that for example
$$
{\bf a}'(s) = ({\bf a}'_+ - {\bf a}'_-)\theta(s) + {\bf a}'_-(s)
\eqno (17)
$$
in a neighbourhood of a kink. (We have chosen the step function to be strict,
$\theta(0)=0$ in order that $|{\bf a}'|\equiv 1$ for all s.) Thus
$$
{\bf a}''(s) = ({\bf a}''_+ - {\bf a}''_-)\theta(s) + {\bf a}''_-(s) +
\delta(s) ({\bf a}'_+ - {\bf a}'_-)
\eqno (18)
$$
Therefore, provided ${\bf b}'.{\bf a}'\neq\pm1$ along $\sigma=\tau$, at least
one of the scalars will be singular at the kink.

To summarize: By considering a general solution to the Nambu action we have
shown that the extrinsic curvature of the worldsheet is bounded at cusps but
singular at kinks. We have therefore confirmed that at kinks the Nambu
approximation breaks down as expected, but obtain the unexpected result that at
cusps the approximation remains valid. This result may have implications for
the particle production from cosmic string loops. In general, one might expect
radiation from a string when the static approximation, i.e.~approximating a
cosmic string vortex by the static vortex solution in a plane orthogonal to the
worldsheet, is no longer a good approximation to the full equations of motion.
In other words, when the string wiggles on scales comparable to its width. We
have shown this is not the case at cusps. Therefore, our result suggests that
any calculations, such as [3], which assume cusp annihilation as a
major source of particle production may be over-estimates.
This means that particle radiation from cosmic strings is
probably closer to its original [7], extremely low,
result.

\vskip 1mm

\noindent{\bf References.}

\nref
M.B.Hindmarsh and T.W.B.Kibble, {\it Cosmic Strings}, hep-ph/9411342

\nref
D.F\"orster, \npb 81 84 1974.

K.Maeda and N.Turok, \plb 202 376 1988.

R.Gregory, \plb 206 199 1988.

P.Orland, \npb 428 221 1994.

M.Sato and S.Yahikozawa, hep-th/9406208.

\nref
R.H.Brandenberger, \npb 293 812 1987.

P.Bhattacharjee, \prd 40 3968 1989.

R.H.Brandenberger, A.Sornborger and M.Trodden, \prd 48 940 1993.

\nref
T.W.B.Kibble and N.Turok, \plb 116 141 1982.

\nref
R.Gregory, \prd 43 520 1991.

\nref
R.Gregory, D.Haws and D.Garfinkle, \prd 42 343 1990.

B.Carter and R.Gregory, hep-th/9410095.

\nref M.Srednicki and S.Theisen, \plb 189 397 1987.

\vskip 8truemm

\end